\documentclass[journal=ascecg,manuscript=article]{achemso}
\usepackage[version=3]{mhchem} 
\usepackage{lineno}

\usepackage{color}
\usepackage{siunitx}
\usepackage{bm}
\usepackage{multirow}
\usepackage{amsfonts}
\usepackage{amssymb} 

\DeclareSIUnit\da{\text{Da}}
\DeclareSIUnit\wtpc{\text{wt\%}}

\author{Jianyi Du$^\P$}
\affiliation[Massachusetts Institute of Technology]
{Hatsopoulos Microfluids Laboratory, Department of Mechanical Engineering, Massachusetts Institute of Technology, Cambridge, MA 02139, United States}

\author{Javier P\'aez$^\P$}
\affiliation[University of Vigo]
{Department of Chemical Engineering, University of Vigo, Vigo, 36210, Spain}

\author{Pablo Otero$^\P$}
\affiliation[University of Vigo]
{Department of Chemical Engineering, University of Vigo, Vigo, 36210, Spain}

\author{Pablo B. Sánchez}
\email{pabsanchez@uvigo.es}
\affiliation[University of Vigo]
{Department of Chemical Engineering, University of Vigo, Vigo, 36210, Spain}

\title[An \textsf{achemso} demo]
{Rapid in-situ quantification of rheo-optic 
	evolution for cellulose spinning in ionic solvents}

\begin{document}



\begin{abstract}

    It is critical to monitor the structural evolution during deformation of complex fluids for the optimization of many manufacturing processes, including textile spinning. However, \textit{in situ} measurements in a textile spinning process suffer from paucity of non-destructive instruments and interpretations of the measured data. In this work, kinetic and rheo-optic properties of a cellulose/ionic liquid solution were measured simultaneously while fibers were regenerated in aqueous media from a miniature wet spinline equipped with a customized polarized microscope. This system enables to control key spinning parameters, while capturing and processing the geometrical and structural information of the spun fiber in a real-time manner. We identified complex flow kinematics of a deformed fiber during the coagulation process via feature tracking methods, and visualized its morphology and birefringent responses before and during regeneration at varying draw ratios and residence time. Meanwhile, a three-dimensional physical rheological model was applied to describe the non-linear viscoelastic behavior in a complex wet-spinning process incorporating both shear and extensional flows. We subsequently compared the birefringent responses of fibers under coagulation with the transient orientation inferred from the rheological model, and identified a superposed structure-optic relationship under varying spinning conditions. Such structural characterizations inferred from the flow dynamics of spinning dopes are readily connected with key mechanical properties of fully-regenerated fibers, thus enabling to predict the spinning performance in a non-destructive protocol.

\end{abstract}

\section{Introduction}

Macromolecular systems undergoing highly non-linear deformation in manufacturing processes are subject to transient evolution of their internal structures, including polymer extension and chain orientations. Such structural evolution on the microscopic level results in temporal and spatial-varying properties at larger lengthscales, which are usually accompanied by significant optical responses and can be captured readily through different microscopic or spectroscopic techniques \cite{klemmNanocellulosesNewFamily2011}. Among all the optical phenomena, birefringent responses arising from flow-induced anisotropy are one of the most accessible optical indicators of 
the structural properties \cite{fullerOpticalRheometryComplex1995a}. For measurements, birefringence visualizes different refraction indices between the ordinary and extraordinary axes and can be visualized using polarized microscopy if the materials are non-opaque \cite{fullerOpticalRheometryComplex1995a, Oldenbourg2003}. Well-established techniques, such as the Berek compensator and its variants have been used to quantify static or slowly-varying birefringence. In contrast, in many manufacturing processes, transient birefringent responses are of critical importance to capture the evolution of internal structures during material deforming or phase change, which is key to the resulting properties. However, a fast and accurate \textit{in situ} measurement of such rheo-optic properties has been addressed in very few occasions to the best of our knowledge \cite{mortimerDeviceOnlineMeasurement1994,mortimerFormationStructureSpinning1996c,Mortimer1996f,mortimerSpinningFibresNmethylmorpholineNoxide1998d}. 

An emerging application that necessitates rapid and accurate monitoring of the transient physical and chemical responses is the 
regeneration of man-made cellulosic fibers (MMCFs) that is aimed to replace conventional cotton fibers with high carbon footprints \cite{maCircularTextilesClosed2019b}. In general, MMCFs are produced by dissolving cellulose and spinning the subsequent solutions in non-solvent media to regenerate fibers. Suitable solvents for mass productions are required to dissolve cellulose with minimal degradation, while producing fibers of high quality and exhibiting operational and environmental benefits \cite{hauruDryJetWetSpinning2014}. As a result, the dissolution and regeneration processes are commonly multi-staged and highly transient with complex disruption and formation of inter-cellulose chain linkages. Such transient dynamics are key to monitor the evolution of the spinning process. A rapid measurement for the temporal evolution of the cellulose crystallinity and its internal structures is closely connected to the performance of the spinning and coagulation stages and optimize the overall process \cite{mortimerSpinningFibresNmethylmorpholineNoxide1998d}. 




Over the past decades, a certain family of compounds named ionic 
liquids (ILs) has proven to dissolve cellulose 
with minimal polymer degradation \cite{swatloskiDissolutionCelloseIonic2002}. ILs consist 
of large ions
with highly delocalized charges \cite{weltonRoomTemperatureIonicLiquids1999}. This 
chemical structure confers them unique physicochemical properties for a wide variety of
applications \cite{miyafujiApplicationIonicLiquids2015c}. Given the huge number of ionic 
combinations, ILs are often referred as solvents with tailored 
properties, which have been described in detail in a number of seminal 
works\cite{weltonRoomTemperatureIonicLiquids1999,Brennecke2001,plechkovaApplicationsIonicLiquids2008d}. In real cellulose processing, the application of ILs is deemed as an 
alternative to the more common Lyocell process to 
produce textile fibers from 
biomass \cite{DeSilva2017,maCircularTextilesClosed2019b, hermanutzProcessingCelluloseUsing2019d}.
In a typical cellulose/IL solution for spinning, (referred as ``spinning dope''), the 
dissolved cellulose, despite losing its network integrity, remains broadly entangled and 
dynamically interactive at high concentrations to retain its spinnability and to enhance 
fiber yield from a spinning process. The spinning performance is a result of the complex 
material evolution; hence, the mechanical and chemical properties of the cellulose/IL 
solutions need to be characterized in a local, real-time manner during the spinning and 
regeneration processes, in which a transient exchange of solvent and anti-solvent media 
occurs to the drawn fibers and progressively reconstruct the cellulose structure. To 
produce high-quality 
cellulose fibers, the spinning parameters need to be optimized based on accurate 
monitoring of the structural information in the process \cite{hauruDryJetWetSpinning2014}.


Conversion of dissolved cellulose into fibers is 
achieved via the wet spinning 
process \cite{hanStudyFiberExtrusion1970a}, in which the dope is 
extruded through a spinneret (diameter $D_0$) at an 
averaged velocity $v_0$ ranging 
approximately from \SI{1}{\metre\per\min} to \SI{5}{\metre\per\min} \cite{michudIoncellFIonicLiquidbased2016c}. Of 
note, the dope often passes
through an air gap (named ''dry-jet wet-spinning'') before entering the coagulation bath filled 
with an anti-solvent media \cite{hauruDryJetWetSpinning2014}. To impose a 
preferred conformation to 
the cellulose chains, a pulling rod named \textit{godet wheel} collects 
the extruded filament at a linear speed $v_g$ higher than 
the extrusion velocity from the spinneret $v_0$, 
where $v_\mathrm{g}/v_0=\Gamma$ is referred as the draw ratio. The spun fibers are 
simultaneously coagulated due to the exchange of solvent and anti-solvent for a 
sufficiently long residence time (RT), during which the cellulose chains link together 
via hydrogen bond formation to regenerate a polymer network. \cite{hauruDryJetWetSpinning2014, sayyedCriticalReviewManufacturing2019e}

Dynamics of the coagulation process are normally 
inferred from the characterizations 
of either the cellulose dopes prior to the spinning 
process, or the fully-coagulated spun fibers in a \textit{post factum} 
manner, and these measurements are retrospectively 
subsumed into a trial-and-error 
process to optimize the spinning design for targeted fiber 
properties \cite{hauruDryJetWetSpinning2014,Mortimer1996f,Kong2005}. The 
knowledge obtained from these measurements is thus statistics-based. As a result, the 
obtained spinning parameters based 
on the implications of specific cellulose-dope systems are not 
necessarily applicable to a wider variety of material and 
spinline configurations. Birefringent responses 
of the fibers in an ongoing regeneration process, in contrast, 
provide an easy and non-destructive probe to the cellulose 
structure, and can be readily connected to 
the resulting fiber dynamics and mechanical responses 
through a rheo-optic relationship. In previous studies, the 
birefringent responses during a fiber spinning process have been briefly captured 
for Lyocell processes 
\cite{mortimerFormationStructureSpinning1996c,mortimerSpinningFibresNmethylmorpholineNoxide1998d} and 
cellulose nanofiber systems \cite{mittalMultiscaleControlNanocellulose2018c}. However, 
the measured 
optical responses are mainly phenomenological and do not readily reveal the underlying 
morphological variations of cellulose chains under 
spinning, largely due to limitations in \textit{in situ} visualization 
tools and rheo-optical interpretations of the constitutive 
parameters extracted from the complex rheology of spinning dopes. The lack of both 
instrumentation and fundamental 
understanding hampers the construction of an accurate structural-property 
relationship, thus delaying an optimal spinning process with 
great industrial potentials. 

To address this limitation, we bridge the gap between \textit{a priori} knowledge of the spinning-dope 
rheology and the structural evolution during the fiber spinning process from directly visualizing 
the birefringent responses of cellulose fibers during coagulation. A customized 
instrument is constructed, comprised of a charge-coupled device (CCD) 
camera and a liquid-crystal (LC) compensator with tunable 
retardation to allow for an accurate and scalable \textit{in situ} measurement of the flow kinematics and birefringent responses of extruded filaments during coagulation. The measured birefringence at varying spinning conditions can be readily connected to the averaged cellulose conformation predicted from the spinning-dope rheology and the corresponding flow kinematics. This relationship allows us to predict the 
mechanical properties of fully-coagulated fibers through simple online observation and independent rheological characterizations of the spinning dope. 
As a case study, we measured the kinematics of fiber spinning and birefringent responses of a selected cellulose/ionic liquid (1-ethyl-3-methylimidazolium acetate) 
system at different spinning configurations and degrees of coagulation. We quantified the temporal evolution of an extruded filament, as well as solvent/anti-solvent exchange during the spinning process. From these measurements, we predicted the morphological evolution in the filament along the spinning direction using a physical constitutive framework based on tube models under complex flow conditions \cite{fullerOpticalRheometryComplex1995a, duAdvancedRheologicalCharacterization2022}. We hereby derive a more comprehensive structure-originated dynamic without performing complex scattering-based structural analysis. Outputs from this study can help accurately predict the evolution of fibers in a more general and scaled-up spinning scenario.

\section{Results}

A miniature spinline is configured with an optical setup 
perpendicular to the direction of fiber drawing for birefringence measurements 
(Figure~\ref{fig:setup_material}a, and real setup in Figure~\ref{fig:setup_material}c). In 
the optical setup, a polarizer and an analyzer are configured on either side of the sample 
with well-positioned angles. A liquid-crystal (LC) retarder is positioned along the 
optical path prior to the measured fiber as a retardance compensator controlled by an 
external circuit for \textit{in situ} measurement (Figure~\ref{fig:setup_material}b). 
Details of the optical setup are presented in the Methods section. Along the fiber 
direction, the spinning dope under shear stress is extruded from the spinneret, and 
subsequently spun under an extensional flow imposed by a faster-spinning godet wheel. Two 
close-up schematics under Figure~\ref{fig:setup_material}a show the fiber kinematics near 
(1) and far from (2) the spinneret, respectively. Near the spinneret, a swollen fiber 
close to the spinneret is expected due to the non-trivial normal stress in the radial 
direction arising from the viscoelasticity of the spinning dope. At a draw ratio 
$\Gamma>1$, the filament undergoes an extension with an imposed accumulated strain of 
$\epsilon=\ln{(v_\mathrm{g}/v_0)}$, during which the cellulose chains reorient towards the 
drawing direction. Far from the spinneret, the fiber has reached the godet wheel velocity 
and moves broadly as a right body in an aqueous coagulation bath as antisolvents. A 
prolonged time period of travel (residence time) in the coagulation bath is provided to 
allow for sufficient exchange of the solvents and anti-solvents, which reconstructs the 
hydrogen bonds between cellulose chains, hereby linking the cellulose chains to form 
stable internal structures. 

In this study, we demonstrated the cellulose regeneration with the prehydrolysis-kraft 
dissolving pulp, \textit{Eucalyptus urograndis}, dissolved in 1-ethyl-3-methylimidazolium 
acetate, \ce{[C2C1Im][OAc]} (Proionic GmbH, Austria), at $c=\SI{5}\%$. This concentration 
of cellulose is selected above its entangled concentration $c_\mathrm{e}$ to 
optimize the fiber yield with minimal amount of solvent needed 
\cite{hawardShearExtensionalRheology2012i,Gericke2009b}. When $c>c_\mathrm{e}$, cellulose chains 
start to entangle and form larger-scale networks, modifying the rheological responses due 
to the deformation and alignment of the collective chain deformation 
\cite{dealyMolecularStructureRheology2006a}. As a result, it is critical to extract the 
flow kinematics during the spinning process and the complex rheology to describe the fiber 
evolution in a spinning process more comprehensively.

\begin{figure}
\centering
    \includegraphics[width= 1.0\textwidth]{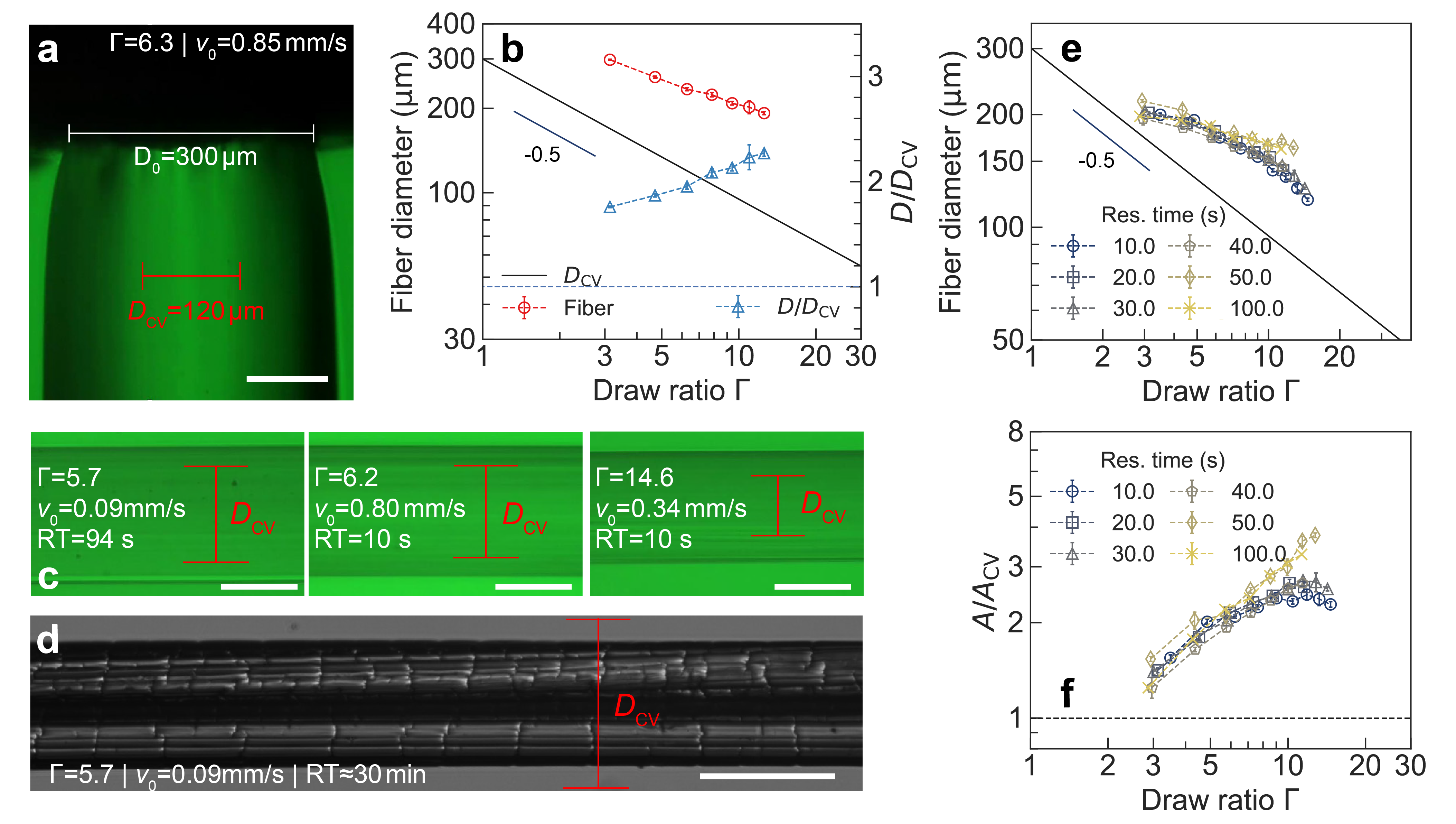}
    \caption{Evolution of fiber geometries in a spinning process. (a) Extruded filament near the spinneret that exhibits die-swelling effects. Draw ratio $\Gamma= 6.3$. Red reference line: $D_\mathrm{CV}$. (b) Fiber diameter $D$ (red circles) and its ratio to that under conservation of volume $D/D_\mathrm{CV}$ (blue triangles) near spinneret ($x= \SI{2}{mm}$) at varying draw ratios. Solid line: Predicted steady-state fiber diameter $D_\mathrm{CV}$ under conservation of volume. (c) Extruded filament sufficiently far from the spinneret ($x=\SI{50}{mm}$) at varying draw ratios and residence times. Red reference line: $D_\mathrm{CV}$. (d) Bright-field image of a fully-coagulated fiber ($\Gamma=\num{5.7}$), 
    showing spatially heterogeneous core-shell structures due to ongoing solvent/anti-solvent exchange. Red reference line: $D_\mathrm{CV}$. (e) Fiber diameter far from the spinneret against draw ratios at varying residence times. The spinneret diameter is fixed at $D_0=\SI{300}{\micro\m}$. (f) Fiber swelling ratio $A/A_\mathrm{CV}$ due to coagulation at varying draw ratios and residence time. Solid line: No change of fiber cross-section ($A=A_\mathrm{CV}$). 
    Scale bars in (a), (c) and (d): \SI{100}{\um}}
    \label{fig:res_dia} 
\end{figure}

\begin{figure}
    \centering
    \includegraphics[width=1.0\textwidth]{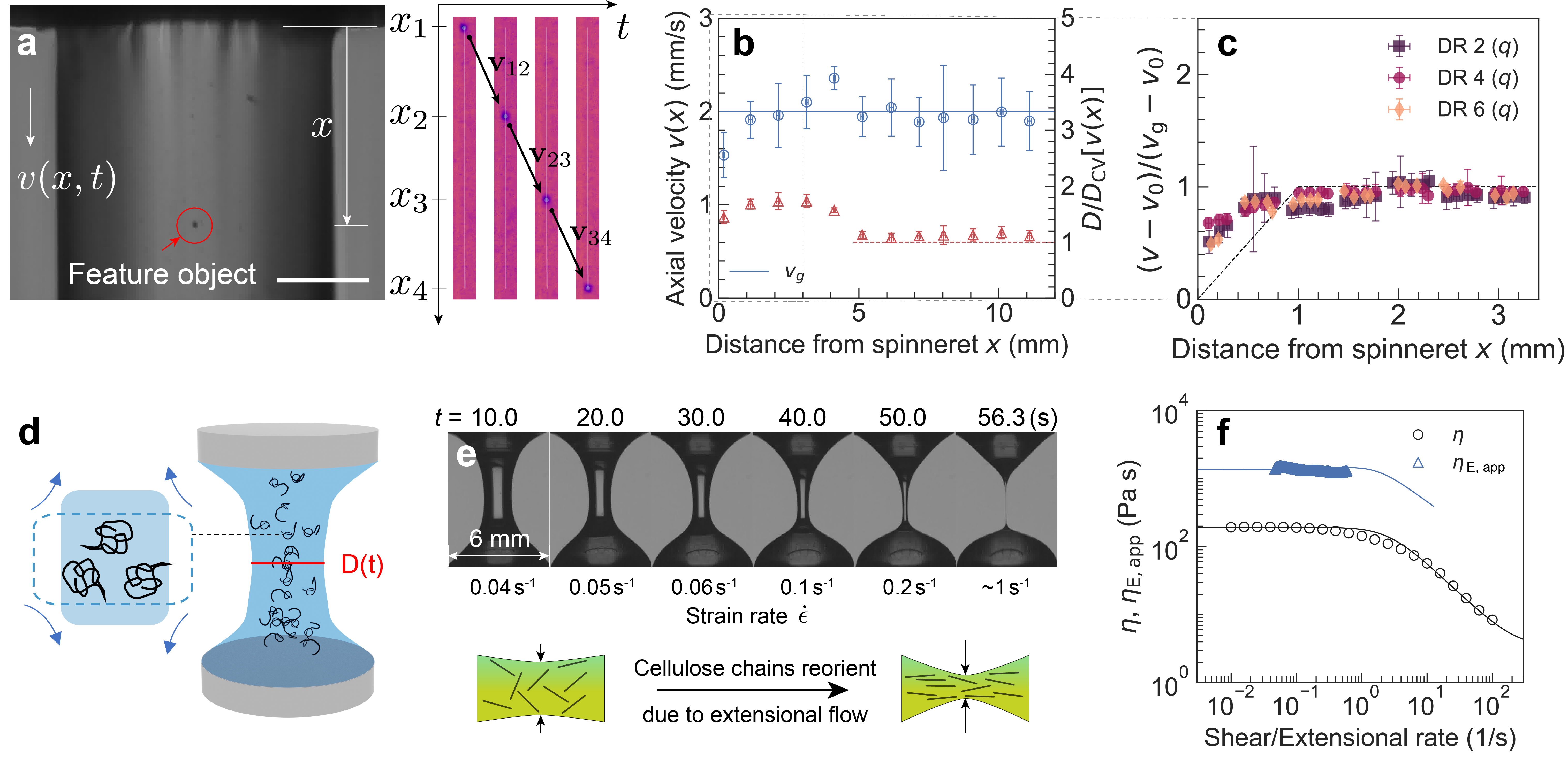}
    \caption{Velocimetry and rheological characterizations of the studied spinning dope. (a) Schematic of feature-tracking velocimetry at varying distance to the spinneret $x$. (b) Blue circles (left $y$-axis): Averaged axial velocity extracted from tracking multiple feature objects (\SI{0}{mm} to \SI{11}{mm}; bin size \SI{1}{mm}) at $\Gamma=\num{2}$. Blue solid line: $v_\mathrm{g}$. Red triangles (right $y$-axis): Ratio of fiber diameter $D$ and the transient fiber diameter under conversation of volume $D_\mathrm{CV}[v(x)]$. Red dashed line: Reference line with no change in fiber volume (c) Normalized axial velocity from averaging the transient velocity measurements (\SI{0}{mm} to \SI{3}{mm}; bin size \SI{0.2}{mm}) with the godet wheel speed ($v_\mathrm{g}$) at varying draw ratios $\Gamma=\numlist{2;4;6}$. Dashed line: approximated kinematic evolution for spun fibers (Equation~\ref{eqn:vel}). (d) Schematic of capillarity-driven breakup extensional rheometry, in which an extensional flow drives polymer-chain deformation. (e) Snapshots of the filament evolution during capillarity-driven thinning. Maximum transient strain rate $\dot{\epsilon}$ is calculated for each snapshot at the necking region. Bottom: Schematic of flow-induced cellulose reorientation under an extensional flow as the strain rate increases. (f) Shear viscosity $\eta$ (black circles) and apparent extensional viscosity $\eta_\mathrm{E,app}$ (blue triangles) plotted against the corresponding strain rates. Black and blue solid lines correspond to the fitting lines from the Rolie-Poly model under the same constitutive parameters. Both shear and extensional rheological characterizations were performed at \SI{80}{\degreeCelsius}.}
    \label{fig:res_vel} 
\end{figure}

Figure~\ref{fig:res_dia} shows the overall filament morphology under varying spinning configurations. In general, we noticed expanded fiber diameter at the spinneret outlet (Figure~\ref{fig:res_dia}a). We measured the fiber diameter at $x=\SI{2}{mm}$, where $x$ is the distance from the end of spinneret along the fiber direction, at different draw ratios imposed by a constant flow rate ($v_0=\SI{0.85}{mm/s}$) from the spinneret but different godet wheel speeds. The extracted fiber diameter $D$ at varying draw ratios (red circles in Figure~\ref{fig:res_dia}b) exhibit a negative power-law trend against the draw ratio, and the values exceed the predicted diameter based on conservation of volume $D_\mathrm{CV}=D_0/\sqrt{\Gamma}$ (solid line in Figure~\ref{fig:res_dia}a and red reference line in Figure~\ref{fig:res_dia}c and d; $D_0=\SI{300}{\micro\m}$), which can be attributed to both the die-swelling effect from the spinning dope that leads to a non-trivial radial normal stress, and the exchange of solvents and antisolvents. The ratio of $D/D_\mathrm{CV}$ (blue triangles) remains above unity and steadily increases with the draw ratios. 
When the fiber travels far from the spinneret, the normal stress component in the radial direction due to die-swelling effect rapidly relaxes (indicated in Figure~\ref{fig:res_vel}b), and the fiber kinematics are progressively dominated by the specified spinning parameters. Figure~\ref{fig:res_dia}c shows the snapshots of fiber morphology at $\Gamma= 5.7, 6.2 and 14.6$ and residence time of \SI{10}{s} and \SI{94}{s}. The measured fiber diameter is significantly larger than the diameter under conservation of volume $D_\mathrm{CV}$ (red reference lines). Consequently, the exchange of solvents and anti-solvents result in a net flow into the filament. In addition to the overall change in the fiber volume, the spatial distribution of different components is radially heterogeneous due to relatively low solvent/anti-solvent diffusivity in the coagulation process. This non-uniformity radial profile can be visualized by observing the fully-coagulated fiber ($\Gamma=\num{5.7}$) under brightfield imaging (Figure~\ref{fig:res_dia}d), in which a core-shell structure can be clearly identified. Similar fiber structures have been characterized in a number of previous studies \cite{Hedlund2017,hauruDryJetWetSpinning2014}. We plotted the fiber diameters at varying draw ratios and resident times (Figure~\ref{fig:res_dia}e). At different residence times, the evolution of fiber diameters broadly overlap and progressively grow beyond the reference diameter $D_\mathrm{CV}$ (black line) as the draw ratio increases. Of special note, the diameter of fully-coagulated fiber significantly decreases below that under conservation of volume (Figure~\ref{fig:res_dia}d). As a result, we justified the change of fiber diameter in a coagulation process primarily attributed from the solvent/anti-solvent exchange. More quantitatively, the swelling ratios $A/A_\mathrm{CV}=(D/D_\mathrm{CV})^2$ can be calculated under the assumption that the fiber swells uniformly along the radial direction (Figure~\ref{fig:res_dia}f). From the figure, the degree of fiber swelling in a coagulation process is largely dominated by the draw ratio. 

From Figure~\ref{fig:res_dia}, the fiber geometries in a spinning process deviate significantly from the predictions under conservation of volume. Consequently, the fiber kinematics cannot be inferred faithfully from the evolution of its diameter. Therefore, we performed feature-tracking velocimetry using the disperse phases in spinning dopes (Figure~\ref{fig:res_vel}a). These features are largely resulted from the partially-dissolved cellulose with a typical size ranging from \SI{10}{\micro\m} to \SI{100}{\micro\m}. While ionic-liquid solvents can dissolve native cellulose at concentrations above \SI{15}\%, the preparation process requires delicate pretreatments at scaled-up production to facilitate dissolution with controllable derivatizing effects \cite{nazariEffectWaterRheology2017e}. As a result, spinning dopes with partially-dissolved cellulose can better represent the material systems used for industrial applications \cite{hermanutzProcessingCelluloseUsing2019d}. We justified that these disperse features can be used for particle-tracking velocimetry (PTV) by calculating the Stokes numbers defined as $\mathrm{St}=\tau_\mathrm{i}/\tau_\mathrm{f}$. Here, $\tau_\mathrm{i}$ is the relaxation time of a feature object in a flow field and is calculated 
from $\tau_\mathrm{i}=\rho_\mathrm{i}d_\mathrm{i}^2/(18\eta)$, where $\rho_\mathrm{i}$ and $d_\mathrm{i}$ are the density and diameter of the feature object, respectively, and $\eta$ is the viscosity of the fluid phase. On the denominator, $\tau_\mathrm{f}=d_\mathrm{i}/U_0$ characterizes the time of flow past the feature object, and $U_0$ is the field velocity. We calculated the Stokes number for a typical spinning process to be $\mathrm{St}=\numrange{1e-10}{1e-8}\ll 1$ based on independent measurements of the material properties, justifying the use of feature objects as tracers. To recover the axial velocity along the spinning direction, we sampled multiple feature objects at different distance $x=\SI{0}{mm} to {11}{mm}$ from the spinneret, and calculated the ''transient'' axial velocity of each feature object from two adjoining frames (Figure~\ref{fig:res_vel}a). The image processing is performed by a third-party computation package \textit{trackpy} \cite{allandanielb.SoftmatterTrackpyTrackpy2021}. We measured the axial velocity at $\Gamma=2$ at different locations and averaged the raw data into specified bin sizes (\SI{1}{mm}) for plot legibility (Figure~\ref{fig:res_vel}b). We noticed the fiber accelerated to the godet wheel speed $v_\mathrm{g}$ (blue solid line) within a short distance ($x\approx\SI{1}{mm}$). A closer look in the range of \SI{0}{mm} to \SI{3}{mm} at varying draw ratios substantiates a consistent ''acceleration length'' $L_0\approx\SI{1}{mm}$ independent of the imposed draw ratios. As a result, the kinematics of extruded spinning dopes in a generic cellulose-fiber spinning process can be approximated as a piece-wise process: When $x<L_0$, the fiber is extended rapidly to reach the desired terminal velocity ($v_\mathrm{g}$), during which an extensional strain of $\epsilon_0=\ln{\Gamma}$ is accumulated. Beyond this acceleration period, the fiber moves broadly as a rigid body and undergoes a coagulation process over an extended time period. We further justified this separable accelerating-coagulation process via the distinct times for acceleration (approximately \SI{1}{s}) and diffusion (approximately \SI{100}{s}), the latter of which is calculated based on the diffusivity measurements during solvent/anti-solvent exchanges from a number of previous studies\cite{nazariEffectWaterRheology2017e,hauruCelluloseRegenerationSpinnability2016d}. Because the kinematics evolve much faster than the diffusion, the deformation of fibers is relatively instantaneous compared with their regeneration through solvent/anti-solvent exchange.  


To understand the dynamics of spinning dopes prior to the onset of regeneration, we 
performed comprehensive rheological characterizations to the spinning dope under both 
shear and extensional flows, which incorporate the deformation in the spinneret and during 
the spinning process, respectively. The shear rheology was measured using a commercial 
rheometer (Physica MCR 101, Anton Paar), and the extensional rheology was characterized 
using a customized capillarity-driven breakup extensional rheometer (CaBER) 
\cite{nazariEffectWaterRheology2017e}. Both measurements were performed at \SI{80}{\degreeCelsius}. The 
CaBER works by rapidly imposing an extensional strain to a fluid 
sample that rests between two coaxial plates, which induces filament pinch-off as a result 
of the driving surface tension and resistance from the material. An apparent extensional 
viscosity can thus be calculated from the measured filament diameter $D(t)$ 
(Figure~\ref{fig:res_vel}d) as $\eta_\mathrm{E,app}=\sigma_\mathrm{sd}/[-\dot{D}(t)]$, where 
$\sigma_\mathrm{sd}$ is the surface tension of the spinning dope (\SI{47}{mN/m} 
\cite{owensUnderstandingDynamicsCellulose2022c,hawardShearExtensionalRheology2012i}). The 
snapshots of the filament show a breakup time of approximately \SI{57}{s}, during which 
the transient strain rates in the necking region of the filament increase from 
0.04 to 1 $s^{-1}$ (Figure~\ref{fig:res_vel}e). In this process, the cellulose 
chains are forced to reorient and the material properties are significantly modified. 
Finally, the extracted shear and apparent extensional viscosities are plotted and compared 
(Figure~\ref{fig:res_vel}f). The spinning dope shows rate-thinning behavior in both shear 
and extensional flows. To extract the structural evolution during the flow deformation, we 
applied a physical constitutive model (Rolie-Poly model) to fit the experimental data in 
both shear and extensional flows following identical procedures in the previous 
study\cite{owensUnderstandingDynamicsCellulose2022c} (solid and dashed lines in 
Figure~\ref{fig:res_vel}f). The fitting lines are in excellent agreement with
the experimental data.

\begin{figure}
\centering
    \includegraphics[width=0.8\textwidth]{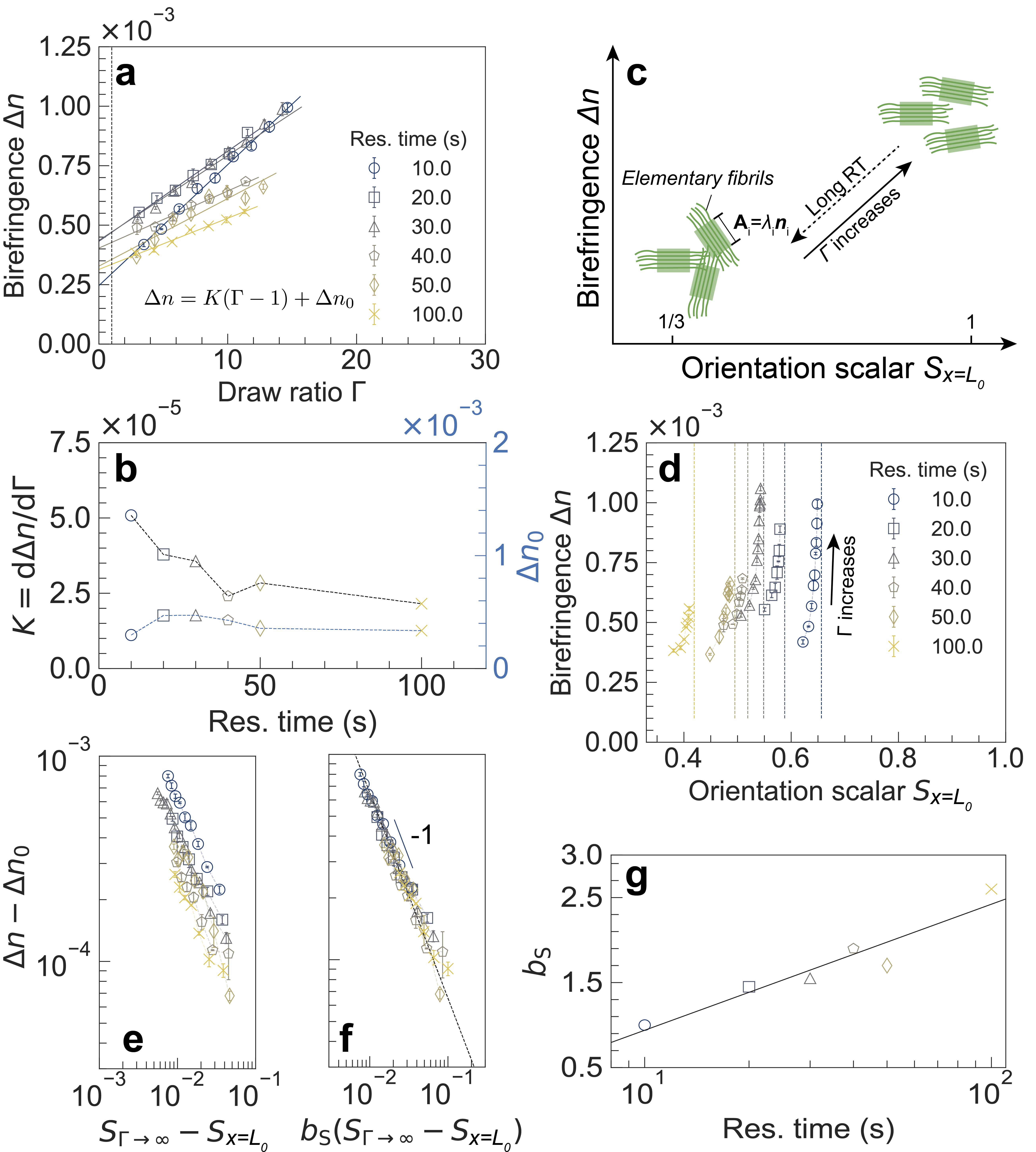}
    \caption{Online birefringence measurement of the fibers during coagulation process and 
    structural interpretations from complex rheology. (a) Birefringence measurements of the 
    fiber in coagulation at varying draw ratios and residence time. Solid lines: Linear 
    fitting lines for each fixed residence time. Dashed vertical line: Draw ratio 
    $\Gamma=1$. (b) Extracted fitting parameters of Equation~\ref{eqn:br_fit} from (a). (c) 
    Illustration of cellulose chain structures at different orientation scalars under 
    varying spinning conditions. (d) Birefringent responses against peak orientation scalar 
    $S(x=L_0)$ at varying residence time. Dashed vertical line: Peak orientation scalar 
    $S_{x=L_0}$ in the limit of $\Gamma\rightarrow\infty$. (e) Birefringent responses due 
    to drawing, $(\Delta n - \Delta n_0)$ against the orientation scalar difference, 
    $(S_\mathrm{\Gamma\rightarrow \infty}-S_{x=L_0})$. (f) Birefringent responses due to 
    drawing against $b_\mathrm{S}(S_\mathrm{\Gamma\rightarrow \infty}-S_{x=L_0})$, pivoted at the 
    residence time of \SI{10}{s}. (f) Extracted horizontal shifting factor $b_\mathrm{S}$ 
    against residence time with the abscissa plotted on logarithmic scale. Solid line: 
    Linear fitting line of the extracted data.} 
    \label{fig:res_br} 
\end{figure}

The predicted structural evolution from the measured spinning dope kinematics and 
rheological responses is subsequently compared with the \textit{in situ} birefringence 
measurements of fibers in a spinning process using an alternating LC retarder. Briefly, 
the LC retarder generates a series of retardation within half the wavelength, which is 
subsequently superposed to the unknown birefringent fibers. The birefringence measurement 
is obtained by extracting the phase difference between the evolution of light intensity 
with or without fibers. Because the phase difference is independent of the overall 
brightness, such birefringence measurements apply semi-opaque material systems as well. 
Using this measuring technique, we found the measured fiber birefringence to increase 
consistently with draw ratios at varying residence times (Figure~\ref{fig:res_br}a), 
showing primary birefringence contributions from fiber extension. We subsequently fit the 
evolution of birefringence with a linear relationship expressed as 
\begin{equation}
\Delta n=K(\Gamma-1)+\Delta n_0,
\label{eqn:br_fit}
\end{equation}
where $\Delta n_0$ is the intercept of birefringence in the absence of extension 
($\Gamma=1$, vertical dashed line). The fitted values (blue squares) remain broadly 
constant, while the slope $K=\mathrm{d}\Delta n/\mathrm{d}\Gamma$ (black triangles) decreases as 
the residence time increases (Figure~\ref{fig:res_br}b). The constant non-zero intercept 
in the absence of extension can be attributed to the non-trivial residue cellulose 
alignment during the flow in spinneret, which remains unaffected during the regeneration 
process. As we attribute the variations in birefringence to the flow-induced anisotropic 
structures resulted from the reorientation of semi-flexible cellulose chains under drawing 
process, the decreased slope $K$ at increased residence time represents enhanced 
resistance to a preferred cellulose alignment under external drawing as the fibers are 
increasingly coagulated, partially due to enhanced fiber stiffness and cellulose 
relaxation. In contrast, the birefringent responses of spun fibers are largely determined 
during the extensional deformation of spinning dopes, parameterized by the draw ratios. 
Figure~\ref{fig:res_vel} has shown that such extensional deformation is imposed in a short 
acceleration length $L_0$ when the majority of the filament remains uncoagulated. As a 
result, the structural properties of the spun fiber at $x=L_0$ can be largely inferred 
from the rheology of spinning dope using previously measured flow kinematics, and become 
an accessible property indicator that readily connects to the fully-coagulated fibers. 
Quantitatively, an orientation tensor $\mathbf{W}$ is commonly used to describe the 
ensemble-averaged orientation of Kuhn steps of all the polymer chains in a solution, and 
has been integrated into a number of constitutive models based on kinetic theories to 
connect micro- and macroscopic properties for polymer solutions and polymer melts 
\cite{mcleishTubeTheoryEntangled2002e,dealyMolecularStructureRheology2006a}. Specifically, 
Owens et al. \cite{owensUnderstandingDynamicsCellulose2022c} has applied the Rolie-Poly 
model to derive a unified mechanical framework to describe the flow behavior of cellulose 
dissolved in ionic liquids over a wide range of strain rates. A frame-invariant scalar can 
be derived from the double-dot product $S=\mathbf{W}:\mathbf{W}$ to describe the 
macroscopic anisotropy arising from preferred orientation of polymer-chain ensembles, 
where $S=1/3$ corresponds to a randomly-oriented distribution, whereas $S=1$ corresponds 
to a well-aligned distribution \cite{likhtmanSimpleConstitutiveEquation2003e} 
(Figure~\ref{fig:res_br}c). 
In a spinning process, the evolution of $\mathbf{W}$ can be calculated from the measured 
flow kinematics in the form of axial velocity $v(x)$. To demonstrate this relationship, we 
approximate the evolution of axial velocity in a coagulation process using a simple linear 
form (dashed line in Figure~\ref{fig:res_vel}c) as
\begin{equation}
    v(x) = 
    \begin{cases} 
        v_0+\dfrac{(v_\mathrm{g}-v_0)x}{L_0}, & x\le L_0 \\ 
        v_\mathrm{g}, & x>L_0 \\
    \end{cases}
    \label{eqn:vel}
\end{equation}
where $L_0$ is the acceleration length measured in Figure~\ref{fig:res_vel}. The 
imposed strain rate during fiber acceleration remains constant and can be calculated as $\dot{\epsilon}(x)=\mathrm{d}v/\mathrm{d}x=(v_\mathrm{g}-v_0)/L_0$ for an extended period of $t_\mathrm{a}=\int_0^{L_0}\mathrm{d}x/v=L_0\ln{(\Gamma)}/(v_\mathrm{g}-v_0)$. The transient extensional rate $\dot{\epsilon}(x)$ can thus be substituted into the 
constitutive equation to calculate the evolution of orientation tensor $\mathbf{W}(x)$. The 
initial condition ($x=0$ in a Eulerian frame) is inferred from the 
steady-state shear flow in the spinneret based on the imposed flow rate and the spinneret geometry. 
By plotting the measured birefringence against a peak orientation scalar $S_{x=L_0}$ defined at $x=L_0$ for each residence time (Figure~\ref{fig:res_br}d), we identified similar increasing trends in the birefringence as the structure becomes more aligned. 
Compared with the imposed spinning parameters, the birefringence provides a more generic 
and consolidated measure for the resulting fiber structure. To show this, we establish a 
superposing relationship between the birefringence and the structural 
parameters under varying spinning conditions. We notice that at a fixed residence time 
(hence $v_\mathrm{g}$), the peak orientation scalar converges to a finite value 
$S_{\Gamma\rightarrow\infty}$ as the draw ratio increases. This finite value can be 
determined by imposing a steady-state extensional rate of $\dot{\epsilon}=v_\mathrm{g}/L_0$, 
which can be shown to produce a mathematically equivalent flow dynamic (dashed vertical 
asymptotes in Figure~\ref{fig:res_br}d). We identified similar asymptotic trends of 
birefringence measurements at different residence time. To render a superposed 
relationship across varying residence times, we replotted the birefringence due to 
spinning, $\Delta n-\Delta n_0$, against 
$S_{\Gamma\rightarrow\infty}-S_{x=L_0}$ (Figure~\ref{fig:res_br}e). We 
found that all curves exhibit a 
power-law decaying trend with a broadly constant power exponent of \num{-1}. A horizontal 
shifting factor $b_\mathrm{S}$ based on the measurement at residence time of \SI{10.0}{s} is 
imposed to further consolidate a master curve at varying residence time 
(Figure~\ref{fig:res_br}f), and the shifting factor $b_\mathrm{S}$ extracted from the least 
square regression exhibits a clear correlation with the residence time 
(Figure~\ref{fig:res_br}g). The last shifting operation is not trivial, because the rate 
of relaxation for cellulose orientation can vary significantly with the residence time, 
and needs to be described separately with an additional superposition. The extracted 
shifting factor $b_\mathrm{S}$ appears to be proportional with the logarithmic residence time 
(black line), which indicates a slow-down in cellulose relaxation as the residence time 
increases. As a result, we justify 
a universal rheo-optic relationship derived from the general constitutive 
model for entangled spinning dopes using the peak orientation scalar. Of special note, the 
birefringence of cellulose fibers is not only function of the draw ratio and the residence 
time, but also of the extrusion speed at the spinneret, which dominates the cellulose 
orientation under rapid extension at the onset of fiber spinning. The mechanical 
properties of fully-coagulated fibers thus vary accordingly, even if the draw ratios are 
identical. 

\begin{figure}
    \centering
    \includegraphics[width=0.8\textwidth]{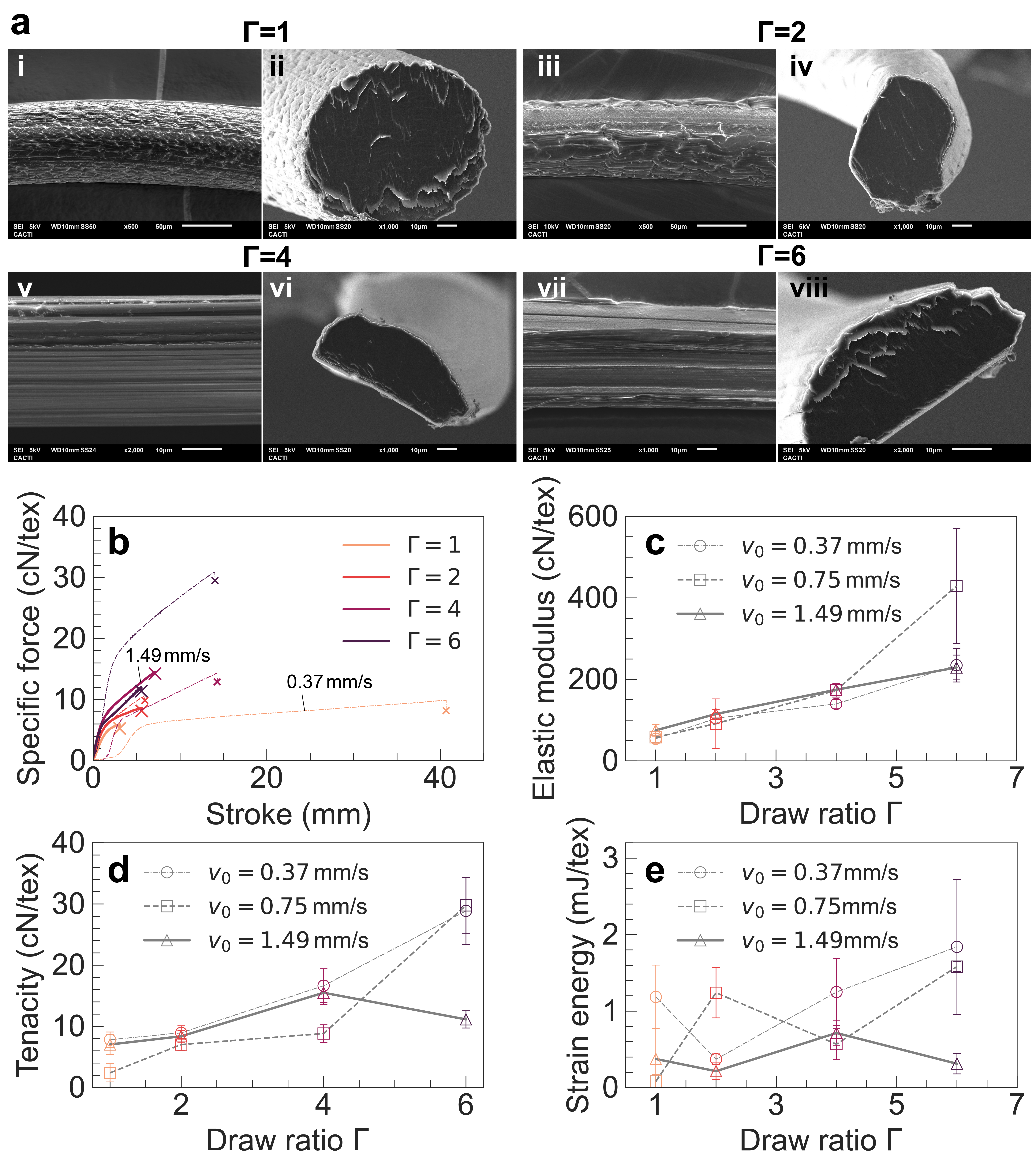}
    \caption{Structural and mechanical characterizations of fully-coagulated 
    fibers. (a) Scanning electron microscopic images of longitudinal (i, iii, v, vii) 
    and cross-sectional (ii, iv, vi, viii) views of the spun fibers at varying draw 
    ratios and extrusion speed of $v_0=\SI{0.75}{mm/s}$. Scale bars are specified in 
    each image. (b) Specific force-stroke curve for fully-coagulated fibers generated 
    at varying draw ratios ($\Gamma= 1, 2, 4, 6$) and extrusion speeds 
    ($v_0=\SI{0.37}{mm/s} to \SI{1.49}{mm/s}$). One dataset under each condition is shown. The 
    points of failure are marked in each curve. (c) Measures of modulus for fully-
    coagulated fibers against draw ratios at varying extrusion speeds 
    ($v_0 = 0.37, 0.75, \SI{1.49}{mm/s}$), extracted from the maximum slope of the 
    specific force-stroke curves. (d) Measures of tenacity for fully-coagulated fibers 
    against draw ratios at varying extrusion speeds ($v_0 = 0.37, 0.75, \SI{1.49}{mm/s}$), 
    extracted from the maximum specific force prior to points of failure. (e) 
    Measures of specific strain energy against draw ratios for fully-coagulated fibers 
    at varying extrusion speeds ($v_0= 0.37, 0.75 and \SI{1.49}{mm/s}$), calculated from 
    integrating the specific force-stroke curve in (b) before points of failure. Three 
    tests were performed for each datapoint in (c), (d) and (e).}
    \label{fig:res_tenacity} 
\end{figure}

The spun fibers under varying spinning condition were collected from the miniature 
spinline, fully coagulated and dried for structural and mechanical characterizations 
(Figure~\ref{fig:res_tenacity}). Longitudinal (i, iii, v, vii) and cross-sectional (ii, 
iv, vi, viii) morphologies at varying draw ratios ($\Gamma=\numlist{1;2;4;6}$ at 
$v_0=\SI{0.75}{mm/s}$) were captured from scanning electron microscopy 
(Figure~\ref{fig:res_tenacity}a; JEOL USA), and stronger extension in the drawing 
direction can be identified as the draw ratio increases. The fiber cross-sections at high 
draw ratios are smoother and progressively deviate from circular shapes, which are 
compatible deformations caused by the horizontal 
rods in the coagulation bath (see experimental setup 
in Figure~\ref{fig:setup_material}) induced 
by the strain in the take-over rod (Godet wheel), demonstrating increased plasticity due 
to enhanced cellulose alignment. 
The mechanical properties of the fully-coagulated fibers were measured 
using a standard mechanical tester (AGS-X STD, Shimadzu) 
equipped with a \SI{10}{N}-load cell according to the standard measuring protocol 
\cite{14:00-17:00ISO50791995}. To substantiate the effects from both draw ratios and 
extrusion speeds on the spinning performance, three extrusion speeds 
($v_0 = 0.37, 0.75, \SI{1.49}{mm/s}$) were tested. Figure~\ref{fig:res_tenacity}b shows 
specific force against stroke at distinct extrusion speeds of \SI{0.37}{mm/s} (thin 
dashed lines) and \SI{1.49}{mm/s} (thick solid lines) at varying draw ratios. We 
identified similar two-stage mechanical responses with elastic and plastic regions under 
varying spinning conditions. However, the magnitudes on both abscissa and ordinate show 
distinct trends. To quantify the mechanical responses of the spun fibers, three specific 
properties were extracted from the force-stroke curve: the stiffness, the tenacity, and 
the strain energy (Figure~\ref{fig:res_tenacity}e). The stiffness, which describes the 
linear elasticity, increases at a higher draw ratio, but remains broadly unchanged at 
varying extrusion speeds (Figure~\ref{fig:res_tenacity}c). Beyond the linear region, the 
tenacity and strain energy, which describe the fiber strength and toughness, respectively, 
increase at a higher draw ratio as well as at a slower extrusion speed.
For an entangled polymer network such as regenerated cellulose fibers, the linear 
elasticity arises from variations in microscopic entropy due to reorientation of cellulose 
chains \cite{dealyMolecularStructureRheology2006a}. As a result, the magnitude of elastic 
moduli is readily connected to the structural conformation after spinning, regardless of 
the transient deformation throughout the process. Figure~\ref{fig:res_br}b, the 
birefringence measurement under no extension ($\Delta n_0$) is broadly constant across 
varying residence times. As result, the structural conformation of the regenerated 
cellulose is largely determined by the draw ratio, hence the linear elastic properties. On 
the contrary, despite a number of studies that have addressed the dependence of tenacity 
on draw ratio \cite{hauruDryJetWetSpinning2014, suzukiAirJetWetSpinningCurdlan2021c}, 
these non-linear mechanical properties appear to be functions of the extrusion speed $v_0$ 
as well. The enhanced tenacity and specific strain energy at lower extrusion speeds have 
been briefly reported previously 
\cite{moriamSpinneretGeometryModulates2021c,liCelluloseFibersCellulose2014b}, which 
attributed the more tenacious and ductile trends in the regenerated fibers at lower 
extrusion speeds to a smaller deformation rate in the spinneret, thus less deformation 
energy in the spinning-dope before spinning and coagulation. However, as stated 
previously, we did not observe significant change in the birefringence of spinning dope 
right after extrusion from the spinneret at $\Gamma=1$ ($\Delta n_0$ in 
Figure~\ref{fig:res_br}b). As a result, we attribute the increased tenacity and toughness 
to the prolonged acceleration time $t_\mathrm{a}$ during the fiber drawing period ($x<L_0$) as 
the extrusion speed decreases at a constant draw ratio. 
By substituting the spinning parameters, we found that such acceleration times under all 
the studied spinning conditions are greater than \SI{0.24}{s}, which remains larger than 
the disengagement time ($\tau_\mathrm{d}\approx \SI{0.21}{s}$) in the Rolie-Poly model 
extracted from the rheological characterizations (Figure~\ref{fig:res_vel}f). As a result, 
despite the fiber drawing induced by a non-trivial draw ratio, which induces significant 
reorientation to the cellulose chains via advection, these chains simultaneously undergo a 
disengaging process and are reorganized to reduce the free energy. This ``annealing-like'' 
process homogenizes the microstructures and gives rise to enhanced resistance to material 
failure at larger external deformation, and is critical to grant strong and tough fibers 
that may find commercial applications.

\section{Conclusions}

In this work, we propose a universal rheo-optic framework to monitor the regeneration 
process for cellulose dissolved in ionic liquids via a wet-spinning. A mini-spinline 
was constructed and integrated with a polarized microscope to visualize the geometry 
and birefringence of spun cellulose fibers in real time at varying draw ratios and 
residence times. Using feature tracking techniques, we identified a broadly constant 
distance within which the fibers are extended upon extrusion from the spinneret. 
Beyond this point, the fibers move in the coagulation bath with minimal deformation 
for an extended period (residence time), where the exchange of solvents and anti-solvents gradually regenerates the cellulose network.

We measured the birefringence of fibers in the spinning process to substantiate the 
microstructural variation. To quantify the flow-induced structural evolution during 
the spinning process, a rheo-optic framework based on the Rolie-Poly model was 
proposed, and the constitutive parameters were extracted from independent shear and 
extensional rheological characterizations. Based on this rheo-optic framework, a 
superposing relationship can be obtained between the optical measurements and the 
inferred structural anisotropy, hence providing accessible indicators of the 
cellulose structures from online birefringence measurements. 

Finally, the mechanical properties of regenerated fibers at varying draw ratios and 
extrusion speeds were measured. While the linear elastic properties appear to be sole 
functions of the draw ratio, we identified enhanced tenacities and strain energies as 
the extrusion speed decreased. We attributed this trend in the non-linear region to 
the lower transient anisotropy of cellulose structures throughout the spinning 
process, which allows for enhanced degree of structural relaxation. As a result, the 
coagulation process is more homogenized to reduce the free energy of formed cellulose 
chains, and to facilitate the growth of larger cellulose networks.


\section{Experimental Section}
\subsection{Material preparations}

The material system applied in this work is prehydrolysis-kraft 
dissolving pulp (\textit{Eucalyptus urograndis}, 93\% 
cellulose I, $M_\mathrm{w}=\SI{269}{\kilo\da}$ 
and $M_\mathrm{n}=\SI{79}{\kilo\da}$ with a polydispersity of \num{3.4}) 
dissolved in 1-ethyl-3- methylimidazolium acetate (\ce{[C2C1Im][OAc]}) provided through courtesy of Prof. Michael Hummel from Aalto University. The 
concentration is selected at \SI{5}{\wtpc} (corresponding 
to $c/c_\mathrm{e} \approx 2.5$, where $c_\mathrm{e}$ is 
the entanglement concentration). Material systems with $c>c_\mathrm{e}$ are selected 
to be consistent with real spinning processes, in which concentrated cellulose 
spinning dopes are generally applied~\cite{hauruDryJetWetSpinning2014}. Spinning dopes 
were prepared following a standard procedure that has been illustrated elsewhere 
\cite{owensUnderstandingDynamicsCellulose2022c}. Briefly, a certain amount of cellulose 
was dissolved at \SI{90}{\degreeCelsius}, in a glass beaker sealed with a PTFE stirrer 
bearing under mechanical mixing for \SI{60}{mins}. After complete dissolution dopes where 
filtered at room temperature through a \SI{7}{\micro \metre}-filter mesh and degassed at 
\SI{70}{\degreeCelsius}.\\

\subsection{Customized spin-line with \textit{in situ} birefringence measurement}

\begin{figure}[htb]
    \centering
    \includegraphics[width=\textwidth]{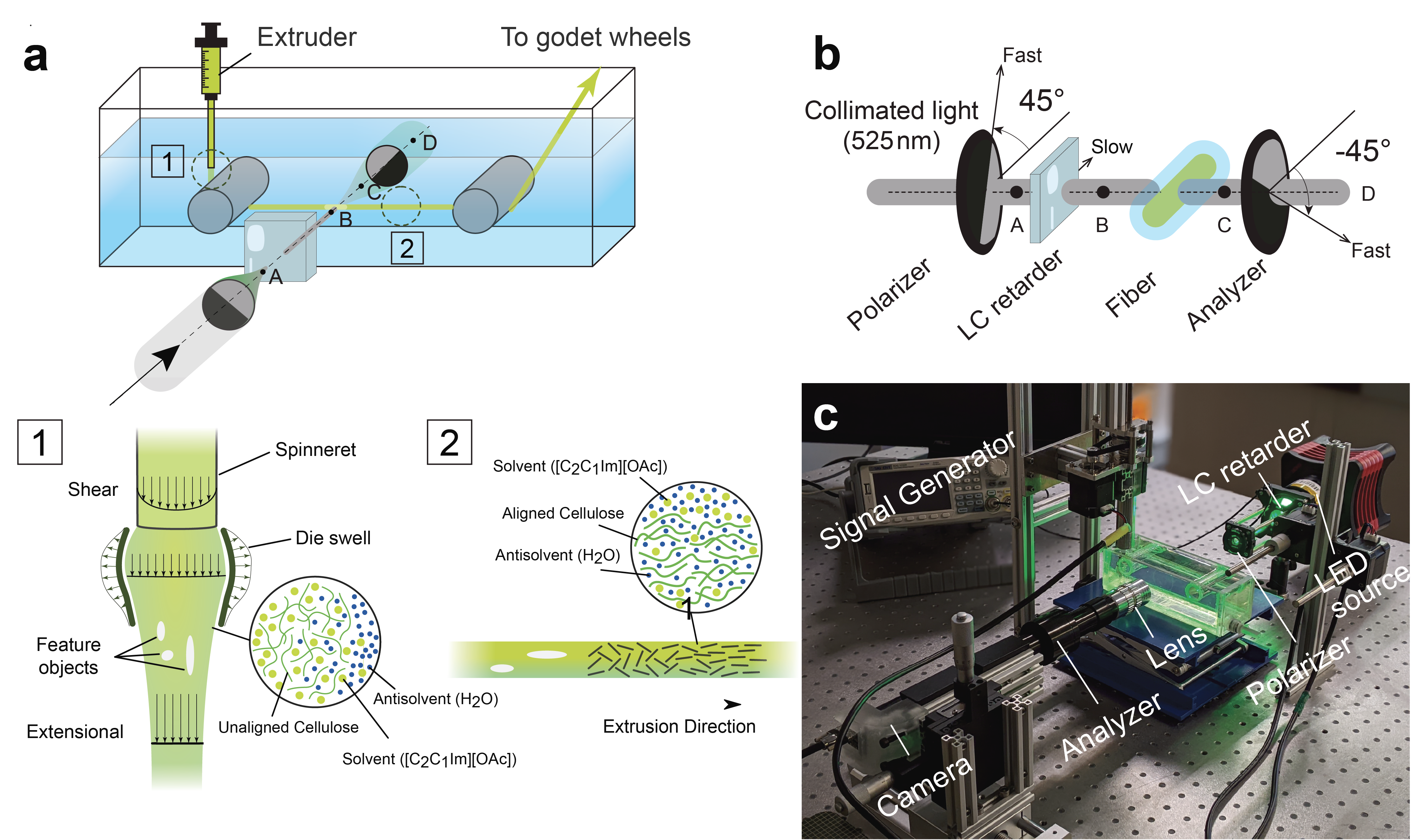}
    \caption{Configuration for \textit{in situ} birefringence measurement on a miniature spinline. (a) Overview of the experimental setup, which constitutes a customized miniature spinline and necessary optical components for polarized microscopy. Schematics underneath show the kinematics and internal structures in the spun fibers (1) near and (2) far from the spinneret. (b) Components the polarized optical setup for birefringence measurement. Position A, B, C and D are consistent with notations in (a). (c) Real experimental setup with primary mechanical and optical components marked in text.}
    \label{fig:setup_material} 
 \end{figure}

 A customized spin-line was constructed (Figure~\ref{fig:setup_material}a), in which the 
 spinning dope is extruded from a custom designed spinneret with the nozzle diameter of 
 \SI{300}{\micro\metre}. The extruded dope undergoes the coagulation process in an aqueous 
 bath as antisolvent between two pillars of an identical size. The fibers, after a fixed 
 time of residence are subsequently reeled with a godet wheel and collected for \textit{post 
 factum} characterizations. During the spinning and coagulation process, the spinning dope 
 undergoes extensional deformation along the spinning direction. As antisolvent diffuses 
 into the dope and induces the gelation of crystalline 
 cellulose, an overall orientation of the 
 cellulose structure is induced. Due to the resulting anisotropy in the overall structure, 
 birefringent responses are generated (Figure~\ref{fig:setup_material}a:1-2), where the 
 slow axis points in the spinning direction. To quantify the birefringent responses, an 
 optical setup (Figure~\ref{fig:setup_material}b) derived from the work of 
 Honda et al. \cite{Honda2018} was constructed. Here, a collimated light (SOLIS-525A, 
 Thorlabs) with a mean 
 wavelength of \SI{525}{\nm} is incident through a 
 polarizer with its slow axis configured at \ang{45} with the horizontal plane. Another 
 analyzer is set with the slow axis set 
 at \ang{-45} against the horizontal plane. Between the polarizer and 
 analyzer, the fiber to be measured is set within the light beam range. To allow for 
 calibrated real-time measurements, a 
 liquid-crystal (LC) retarder is installed long the optical path with 
 the slow axis set at \ang{0}.

\begin{figure}[!htb]
    \centering
    \includegraphics[width=\textwidth]{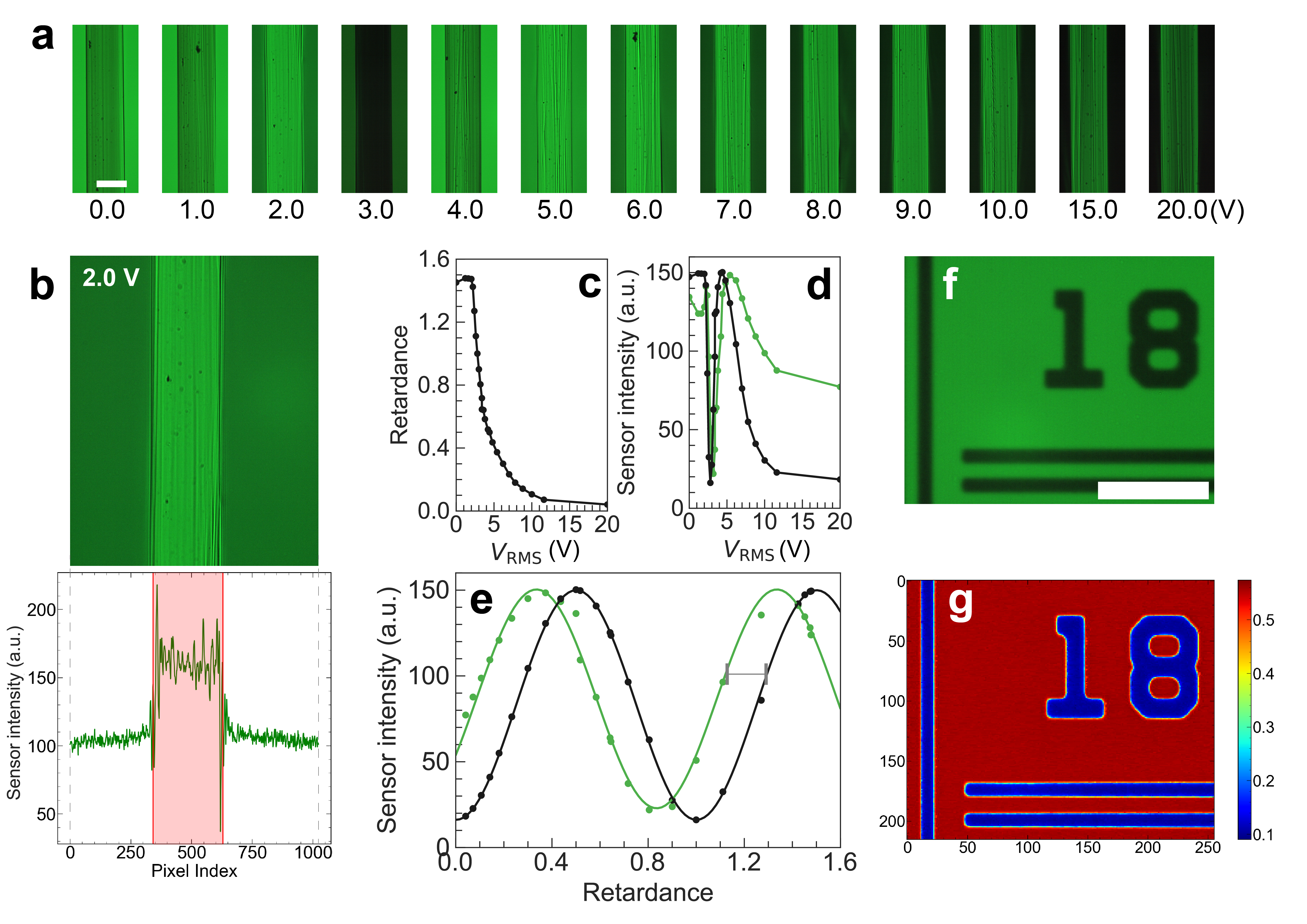}
    \caption{Method overview for fiber detection and birefringence measurement at 
    $\Gamma=11.4$ and residence time of \SI{39}{s}. The wavelength of monotonic light 
    source is \SI{525}{nm}. (a) Snapshots of extruded fibers captured at different root 
    mean square (RMS) voltage levels from the LC retarder. (b) Edge detection for extruded 
    fibers with non-smooth surfaces. Top: Snapshot of the extruded fiber at voltage of 
    \SI{2}{V} from the LC retarder. Bottom: Extracted sensor intensity along cross-
    sectional direction averaged in the longitudinal direction. (c) Calibration curve of 
    retardance for the applied LC retarder. (d) Measured image sensor intensity for the 
    extruded fiber (green) and the background (black) at different RMS voltages. (e) 
    Measured image sensor intensity for the fiber (green) and the background (black) at 
    varying LC retardance. (f) Test birefringent target at $V_\mathrm{RMS}=\SI{0}{V}$. (g) 
    Extracted retardance from the test birefrigent target in (f). Scalebars in (a) and (f) 
    are \SI{100}{\micro\m}.}
    \label{fig:processing} 
\end{figure}

In a birefringence measurement, a well-modulated voltage profile is provided to the LC 
retarder, which induces a temporally-evolving predetermined birefringence along the 
optical path. Such birefringence induced by the LC-retarder superposed with the intrinsic 
birefringence from the sample results in alternating bright-dark image snapshots at 
different root mean square (RMS) voltage levels (Figure~\ref{fig:processing}a). The net 
optical retardance from the fiber can be determined by subtracting the measured retardance 
in the background from that at the fiber centerline. The fiber birefringence can be 
subsequently determined given the fiber geometry. We implemented a simple scheme to 
determine the fiber edges by locating the maximal image gradient along a cross-sectional 
cut line (Figure~\ref{fig:processing}b). Subsequently, the fiber centerline and the 
background regions can be identified. Of note, some LC retarder voltage values are 
insufficient to generated highly-contrasted background-fiber interface, leading to an 
incorrect estimation of the fiber diameter. To correct such miscalculation, the recorded 
measurement for fiber diameter is taken as the median measurement over a cycle of voltage 
iteration from the LC retarder. 

For simplified calculation, we assumed negligible reflectance and absorbance for the 
polarizers. Given the previously-stated slow-axis orientations for all the optical 
components, the resulting light transmittance (dimensionless) is given by 
Equation~\ref{eqn:transmittance} as
\begin{equation}
    \label{eqn:transmittance}
    T_{\Delta\phi}= \dfrac{I_\mathrm{\Delta\phi}}{I_0}=\sin^2(\pi\Delta \phi)=\sin^2(\pi\Delta n d/\lambda),
\end{equation}
where $\Delta \phi=\phi_\mathrm{f}-\phi_\mathrm{LC}=(\Delta n_\mathrm{f}-\Delta n_\mathrm{LC}) D/\lambda$ 
is the dimensionless retardance difference between the fiber sample and the LC retarder at 
an incident light wavelength $\lambda$. The birefringence of the fiber sample and the LC 
retarder are $\Delta n_\mathrm{f}$ and $\Delta n_\mathrm{LC}$, with the optical path lengths 
identical to the fiber diameter $D$ and the LC retarder thickness, respectively.

From the LC retarder, its retardance $\phi_\mathrm{LC}(V_\mathrm{RMS})$ is a function of the input 
voltage, $V_\mathrm{RMS}$ (Figure~\ref{fig:processing}c, provided by the manufacturer). We 
note that in the accessible range of the retardance, the transmittance of the LC retarder 
$T_{\Delta\phi}$ is non-monotonic regarding to $V_\mathrm{RMS}$ (black line in 
Figure~\ref{fig:processing}d), as well as when superposed with the birefringent fiber 
(green line in Figure~\ref{fig:processing}d). The two measured image intensities are 
subsequently plotted against retardance using an interpolated form of the retardance 
calibration curve (Figure~\ref{fig:processing}e; black: LC retarder; green: LC retarder 
superposed with fiber; gray scale: phase difference). The two image intensity responses 
exhibit periodical patterns with a non-trivial phase difference. From 
Equation~\ref{eqn:transmittance}, the light intensity follows a sinusoidal form regarding to 
the LC retardance $\phi_\mathrm{LC}$ as 
\begin{equation}
I_{\Delta \phi}(\phi_\mathrm{LC})=\dfrac{1-\cos[2\pi(\phi_\mathrm{f}-\phi_\mathrm{LC})]}{2}I_0.
\end{equation}
In practice, the numerical intensity extracted from the image pixels slightly deviates 
from a sinusoidal form due to its non-linear correlation with the light intensity. 
Nevertheless, a phase difference between the two periodical patterns can still be 
identified via numerical fitting of a sinusoidal function $f(\phi)=A\sin[2\pi(\phi+B)]+C$, 
where $A$, $B$ and $C$ are the fitting parameters. Finally, the difference in the values 
of $B$ coincides with the retardance of the fiber.


\medskip
\textbf{Acknowledgements} \par 
P.O. and P.B.S. were supported by Ministerio de Ciencia e Innovaci\'on under the grants 
PRE2020-093158 and RYC2021-033826-I respectively. J.D. and P.B.S thank Crystal Owens and Prof. Gareth H. McKinley from MIT for the insightful discussions.

\textbf{Conflict of Interest} \par 
The authors declare no conflict of interest. 

\textbf{Data Availability Statement} \par 
The data that support the findings of this study are available from the corresponding author upon reasonable request.

\medskip

%

\textbf{References}\\
\bibliographystyle{unsrt}
\bibliography{manuscript} 



\end{document}